# The representation of asteroid shapes: a test for the inversion of Gaia photometry

A. Carbognani [a, b, 1], P. Tanga [b], A. Cellino [c], M. Delbo [b], S. Mottola [d], E. Marchese [b]

[a] Astronomical Observatory of the Autonomous Region of the Aosta Valley (OAVdA), Italy
[b] Astronomical Observatory of the Côte d'Azur (OCA), France
[c] INAF, Astronomical Observatory of Torino (OATo), Italy
[d] DLR, Institute of Planetary Research, Berlin, Germany

**Abstract**
It is common practice nowadays to derive spins and 3D shapes of asteroids from the inversion of photometric light curves. However, this method requires, in general, a good number of photometric points and dedicated observing sessions. On the other hand, the photometric observations carried out by the Gaia mission will be sparse and their number relatively small.
For this reason, a multi-parametric shape described by a large number of elementary facets cannot probably be derived from Gaia data alone. Therefore, the Data Processing and Analysis Consortium (DPAC), implemented a simpler solution as an unattended data reduction pipeline which relies on three axial ellipsoids for the shape representation.
However, overall accuracy of such triaxial shape solutions has to be assessed. How adequate is an ellipsoidal approximation to represent the overall properties of an irregular body? Which error is made on the volume in comparison a more accurate model with irregular shape?
To answer these questions, we have implemented numerical procedures for comparing ellipsoids to more complex and irregular shapes, and we performed a full simulation of the photometric signal from these objects, using both shape representations. Implementing the same software algorithm that will be used for the analysis of Gaia asteroid photometry, rotation period, spin axis orientation and ellipsoidal shape were derived from simulated observations of selected Main Belt asteroids assuming a geometric scattering model (work is in progress for more complex scattering models).
Finally, these simulated Gaia results were compared to check the relevance of the ellipsoidal solution in comparison to multi-parametric shapes. We found that the ellipsoids by photometry inversion are closely similar to the best-fitting ellipsoids of the simulated complex shapes and that the error on the asteroid volume (relative to a complex shape) is generally low, usually around 10%.

Keywords: Asteroids photometry, Asteroids volume, Gaia mission.

## 1. Introduction: photometry, spin and shapes

Photometry has been one of the first observing techniques adopted to derive information about the physical properties of asteroids. As of February 2011, there are nearly 4200 reliable rotation periods listed in the Asteroid Light Curve Data Base[2], maintained by Alan W. Harris, Brian D. Warner and

---

Petr Pravec (Warner et al., 2009). As we can see, the number of known periods is still low when compared with the hundreds of thousands of known asteroids (about 550,000 to date). A rotation period can be ideally obtained from the analysis of a single light curve. If several light curves are available, obtained at different apparitions, it is also possible to determine the sky orientation of the spin axis and the shape of the object (Kaasalainen et al., 1992).

If absolute dimensions and mass are known, the shape can be used to estimate the mean (or bulk) density, a crucial physical parameter related to the internal structure of the body. Absolute dimensions and albedos of asteroids are, in general, derived through the so-called radiometric method, though for larger asteroids the diameter can be measured also directly. Thermal radiometry makes use of measurements of the thermal infrared radiation that the object emits at mid-IR wavelengths (5–20 μm) and of its visible reflected light, combined with a suitable model of the surface thermal emission (Harris and Lagerros, 2002).

The European Space Agency's Gaia Mission (launch planned around March 2013), although not explicitly designed to study Solar System's minor bodies, will supply a large amount of sparse photometric data on about 300,000 asteroids down to a visible magnitude of +20 (Mignard et al., 2007; Hestroffer et al., 2010). Each object will be observed up to about 80 times, in a variety of observing circumstances and with this scenario (i.e. with so few data), deriving rotational and shape properties from photometric data is a challenging problem. For this reason, inversion of Gaia asteroid photometry will be made assuming that the objects have a simple three-axial ellipsoid shape. The Gaia data will provide an excellent opportunity to expand the database of the known basic physical properties of asteroids, but how accurate will the approximation be? This is the main issue we would like to address in this paper.

## 2. Simulation of Gaia observations and data processing

In order to provide an answer to the previous question, we simulated the whole observational process, from the generation of synthetic photometric fluxes to the light curve computation, by a dedicated pipeline called "Runvisual" (developed in the C-language), specifically implemented to assess the expected performance of the asteroid photometry inversion. For the generation of fluxes, we described the objects as multi-parametric 3D-shapes represented by triangular elementary facets. In the following we refer to them as "complex models". The asteroid complex models are used to:

(a) Compute the best-fit ellipsoidal models of the assumed complex shapes (see later)
(b) Generate Gaia synthetic photometric observations.

So far, we have used the complex models of eight S-type Main Belt Asteroids (MBAs) obtained by the numeric inversion of ground-based light curves (Kaasalainen et al., 1992; Durech et al., 2010): 3 Juno (effective diameter 234 km), 532 Herculina (222 km), 9 Metis (190 km), 192 Nausikaa (103 km), 584 Semiramis (54 km), 484 Pittsburghia (32 km), 1088 Mitaka (9 km) and 1270 Datura (6 km), corresponding to increasing irregularity in shape and decreasing effective diameter. Note that all the selected objects have a unique solution for the spin axis. The complex models (all having convex shapes) of the MBAs were obtained from the Database of Asteroid Models from Inversion Techniques (DAMIT; Duerch et al., 2010). The database and its web interface[3] are operated by The Astronomical Institute of the Charles University in Prague, Czech Republic.

---
[3] http://astro.troja.mff.cuni.cz/projects/asteroids3D/web.php

The computation of a best-ellipsoid fit (i.e. point (a)) is a process taking place in two steps.

The first step consists of the calculation of the major axis and of the intermediate axis of the ellipsoid in the asteroid X-Y plane as best-fit ellipse (see Fig.1 for an example). To this purpose we compute the average points from the vertices points of the asteroid profile (i.e. the points that fall within a small distance from the X-Y plane in a certain azimuth interval), in order to avoid accumulation of points along the profile in the plane. In this way, no regions of the asteroid weigh more than others and none could alter the position of the best-fit ellipse. With the angular averaged data we can compute the mean of the X and Y coordinates that can be assumed as a guess of the center of the ellipse. With the same averaged points, a guess is made of the semi-major, semi-minor axis and inclination of the ellipse. Finally we find the ellipse parameters for which the Root Mean Square (RMS) is minimal. The RMS is the algebraic distance of the observed points from the equation of the canonical ellipse.

In the second step we compute the third ellipsoid axis along the Z-axis by imposing that the volumes derived from both the complex and the ellipsoidal shape are equivalent. In this way, by construction, the equatorial plane and the spin of the best-ellipsoidal shape are the same as the complex shape. The best-ellipsoidal shape thus obtained is taken as reference for comparison with the ellipsoid obtained by inversion of the simulated Gaia asteroid photometry.

The dates of Gaia observations have been simulated using software written by F. Mignard and P. Tanga and implemented in the Java computer language by Christophe Ordenovic (Observatoire de la Côte d'Azur). The software simulates the Gaia observation sequence for any Solar System object, providing for each observation the corresponding Gaia-centric and heliocentric distance and phase angle.

Apparent magnitudes (i.e. point (b)), were computed at simulated observation epochs for the chosen MBAs, using their (known) spin, period and corresponding complex models. The right orientation of a complex model at epoch $t$ is given by a transformation between vectors $r_{ast}$ in the asteroid co-rotating coordinate frame and vectors $r_{ecl}$ in the ecliptic coordinate frame using appropriate rotation matrices. Light scattering effects on asteroid surfaces were modeled using both purely geometric scattering(i.e. assuming a scattered flux proportional to the illuminated asteroid cross section as seen by the observer) and the scattering model adopted by Kaasalainen et al. (2001), a combination of 10% of Lambert scattering and 90% Lommel-Seeliger. Because the analysis of the simulations with the latter scattering law is still in progress, we will only report the early results obtained with the geometric scattering.

In order to simulate the Gaia photometric observations, "Runvisual" reads from an ephemeris file the observation epoch and the asteroid's heliocentric and geocentric coordinates, rotates the asteroid model according to the Sun-Gaia-object geometry and computes the normal vector to the asteroid facets. At this stage, only the facets illuminated by the Sun and seen from Gaia are selected and, by applying the appropriate scattering model, a magnitude value is computed. For an example of the simulated magnitudes see Fig. 2.

The simulated Gaia observations obtained from "Runvisual" were inverted by using the "genetic" algorithm (Cellino et al., 2009) written for Gaia data processing, obtaining rotation period, pole coordinates, ellipsoidal shape (b/a, c/a), and phase-mag slope for each simulated set of observations. The results of the inversion were then compared with the input shape. One of the most interesting indications about the accuracy of the results are obtained by comparison with the best-fit ellipsoidal model derived from the complex shape. As a matter of fact, in this way we can verify the physical quality of the ellipsoids obtained by using the genetic algorithm. In our simulation of the Gaia photometric observations we didn't study the effect of photometric noise on the results of the light curve inversion. This aspect will be addressed in a more complete work. However, in Cellino et al. (2009), a first analysis of the effect of the noise and of the number of observations indicates that random errors up to 0.03 mag do not invalidate the results of the inversion method.

In general, the values of the rotation periods of the genetic ellipsoids are identical to those of the complex models (the rotation period was precise within $10^{-4}$-$10^{-5}$ hours), the only exception is the period of 3 Juno, which result twice of the real value. Our results suggest also that the "genetically derived" ellipsoids found by photometry inversion closely resemble the best-fitting ellipsoids of the complex shapes. The axial ratios between the genetic inversion and the best-ellipsoidal models (in geometric scattering model) are b/a = 0.94 ± 0.06 and c/a = 0.95 ± 0.09, while the spin coordinates difference, in ecliptic longitude and latitude, between the genetic inversion and the complex or best-ellipsoidal models are Δλ = 2 ± 1° and Δβ = 3 ± 8° (see Fig. 3 and 4). The only great exception is for the spin longitude of 584 Semiramis, where there is a difference of about 180° compared to the real value. In any case the spin solution is unique, i.e. there are no cases in which two spin values are compatible with the observations. We note also that, according to Torppa et al. (2008), the overall difference (RMS) between complex and best-ellipsoidal model is not very important for a good pole fit, confirming similar results by Cellino et al., (2009).

### 3. Estimate of the error on the volume of the asteroids

In what follows we estimate the error on the asteroid volume using the ellipsoidal approximation from the Gaia genetic inversion with respect to the complex shape from DAMIT. It should be noted that the volume of the complex shape is not the true asteroid volume, because there are other uncertainties, namely the effect of concavities, to which the full inversion method is insensitive, and the overall scale of the body, as determined usually by thermal radiometry. Indeed, if we compare the radiometric diameters from simple thermal model with diameters from occultation measurements for Main-Belt asteroids, the uncertainty generally is about 10%, e.g. 30% on the volume (Harris and Lagerros, 2002). For a more detailed discussion on the uncertainties of the diameter of asteroids see Carbognani, (2011).

Once known, the true volume and mass of an asteroid, make it possible to derive the bulk density, a crucial parameter for the analysis of the internal structure of the body. Gaia will only make it possible to measure the mass of the 150 largest asteroids, thanks to mutual close approaches and measurement of the consequent gravitational perturbations of the orbits (Mignard et al., 2007).

The genetic algorithm determines the best-fit ellipsoid using the geometric albedo of the asteroid under the condition that the average magnitude of the complex model is the same. With this condition, the surface extension of the genetic ellipsoid and of the complex shape are the same in good approximation. The surface of the complex shapes, denoted by $S_{true}$, can be computed analytically as a sum of modules of vector products while, unlike the surface area of a sphere, the surface area of a general ellipsoid cannot be expressed exactly by an elementary function.

An approximate formula for the surface area of scalene or triaxial ellipsoid ($a_g > b_g > c_g$) is the following (Knud Thomsen's formula):

$$S_g \approx 4\pi a_g^2 \left( \frac{\left(\frac{b_g}{a_g}\right)^p + \left(\frac{c_g}{a_g}\right)^p + \left(\frac{b_g}{a_g}\right)^p \left(\frac{c_g}{a_g}\right)^p}{3} \right)^{1/p} \qquad (1)$$

With $p=1.6075$. Requiring that $S_g = S_{true}$, from equation (1) we can derive the semi-major axis $a_g$ of the genetic ellipsoid (the ratio between the semiaxes are known by genetic inversion). Finally, the absolute relative error on the volume is given by:

$$\frac{\Delta V}{V_{true}} = \left| \frac{V_{true} - V_g}{V_{true}} \right| = \left| 1 - \frac{V_g}{V_{true}} \right| \qquad (2)$$

As in the case of the surface, the volume $V_{true}$ of the complex model can be computed analytically, while the volume $V_g$ for the genetic ellipsoid is given by the simple formula: $V_g = \frac{4}{3}\pi a_g b_g c_g$.

The results from equation (2), in the case of geometric scattering, are shown in Fig. 5. As we can see, the volume errors are not greater than 16% (the worst case is that of 484 Pittsburghia) and there is only a weak correlation with the effective diameter (the probability of linear correlation is about 5.2%). While the effective diameter increases, the volume error tends to decrease as might be expected, as a result of the decreasing complexity of the shape. Of course this error on the volume of a complex model represents a lower limit of the error on the true volume of an asteroid, since it does not take into account possible concavities. We can conclude that the triaxial approximation is a minor sin in the face of the other more dominant uncertainties, as radiometric diameter, in estimating the real volume of an asteroid.

**4. Conclusions**

We have confirmed that, in the geometric scattering approximation, asteroid rotation periods can be determined with high accuracy by means of Gaia data, that the solution for the spin is unique and that pole coordinates and periods derived from inversion are not strongly sensitive to the Root Mean Square between complex and best-ellipsoidal model. Moreover, our findings suggest that the "genetically derived" ellipsoids found by photometry inversion are closely similar to the best-fitting ellipsoids of the complex shapes and that the error on the asteroid volume using genetic ellipsoids is low, usually around 10%.

Of course, we are aware that we are considering here simulated data, only. On the other hand, a previous application of the inversion method to a set of real photometric data obtained by the Hipparcos satellite (Cellino et al., 2009) has already shown that the adopted algorithm is able to give correct inversion in a large variety of situations, even when the data are known to be affected by significant errors.

Bearing in mind these limitation, our conclusions are that the representation of asteroids as genetic ellipsoids does not heavily affect the physical parameters that can be derived (rotation period, spin), and that it remains sufficiently close to the real forms, although unable - in many cases - to reproduce the details of single light curves. However, research on individual cases is probably needed to identify those cases that can present remarkable discrepancies. Their identification should be possible on the basis of light curve properties. The next step for this work is to extend these results to more realistic scattering models and to asteroids with low equatorial elongation, to check more thoroughly the robustness of the Gaia solution. Also the expected errors in the Gaia photometric data have also to be taken into account.

**Acknowledgments**

We thank Alan W. Harris and an anonymous reviewer for their helpful comments.

**Figures**

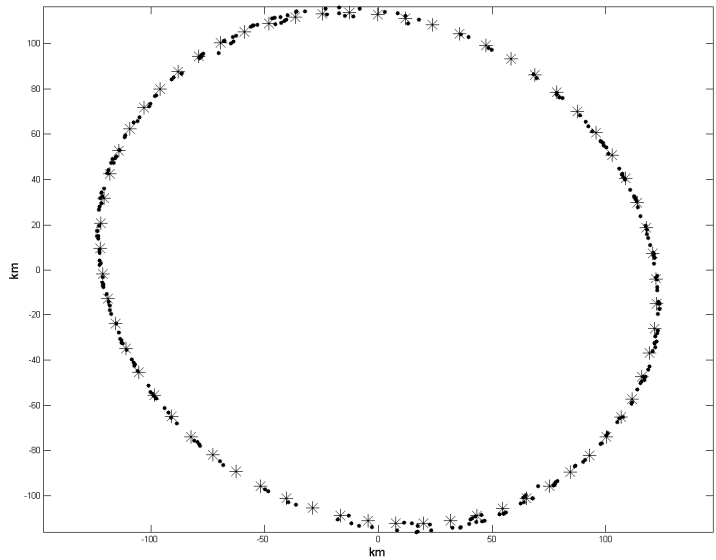

**Figure 1** – Equatorial plane of asteroid 3 Juno. The black dots are the projection of the vertices of the complex model with quote under 0.2 of the Z coordinate, while the asterisks indicate the best-fit ellipsoid.

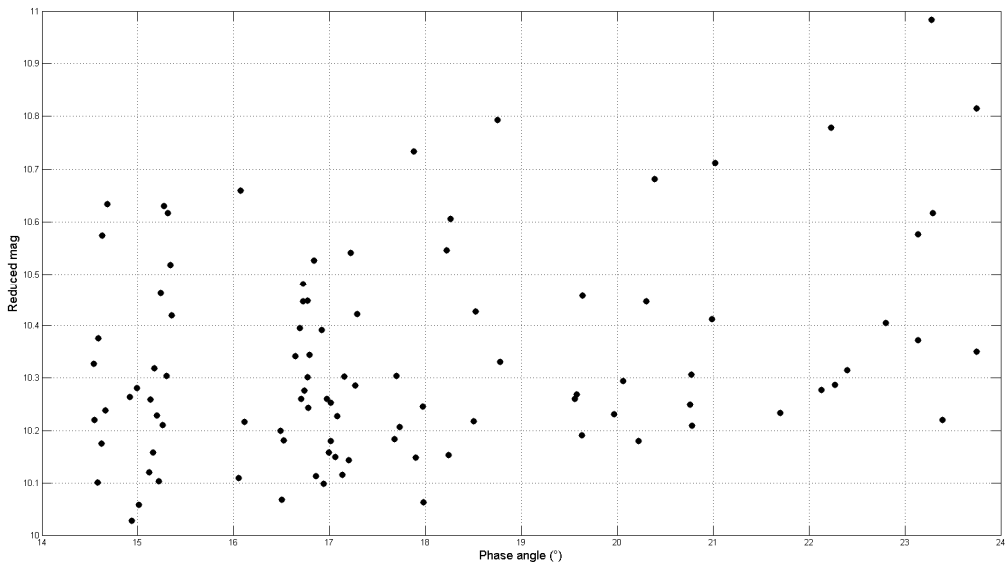

**Figure 2** - The simulated photometric plot, e.g. reduced magnitude vs. phase angle, for the asteroid 484 Pittsburghia obtained with "Runvisual".

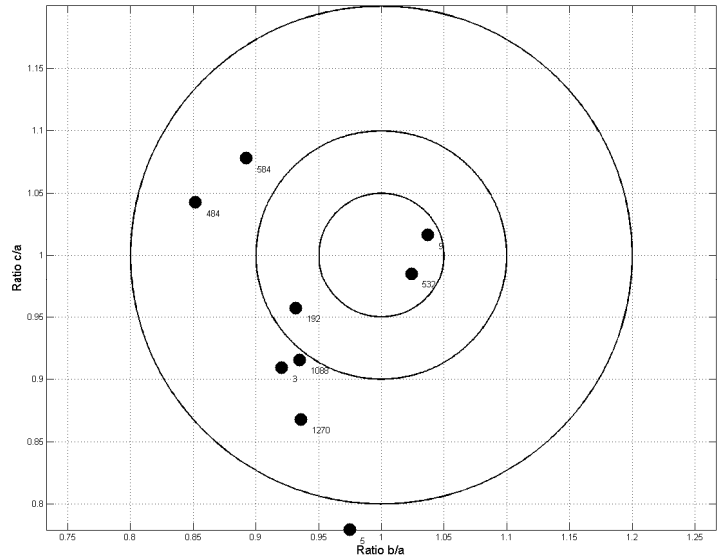

**Figure 3** - Ratios between the axial ratios c/a and b/a from genetic inversion and the corresponding values from the best-ellipsoidal models (geometric scattering model). The asteroids are identified with their progressive number.

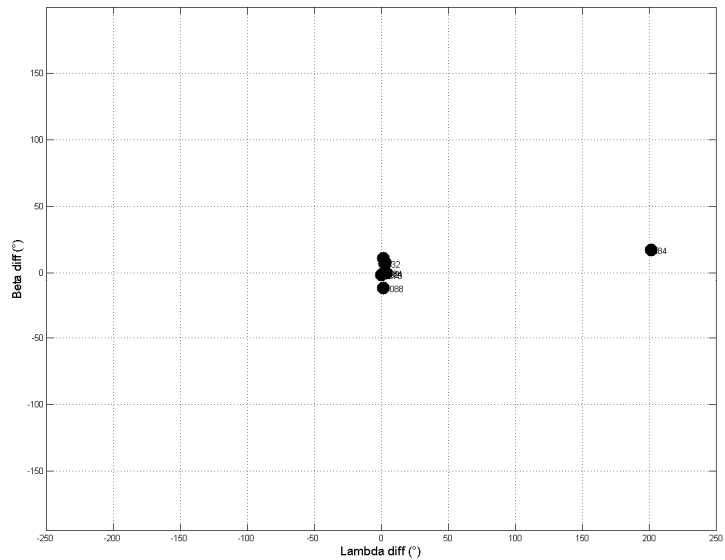

**Figure 4** - Spin coordinates difference between the genetic inversion and the complex or best-ellipsoidal models (geometric scattering model). The asteroids are identified with their progressive number.

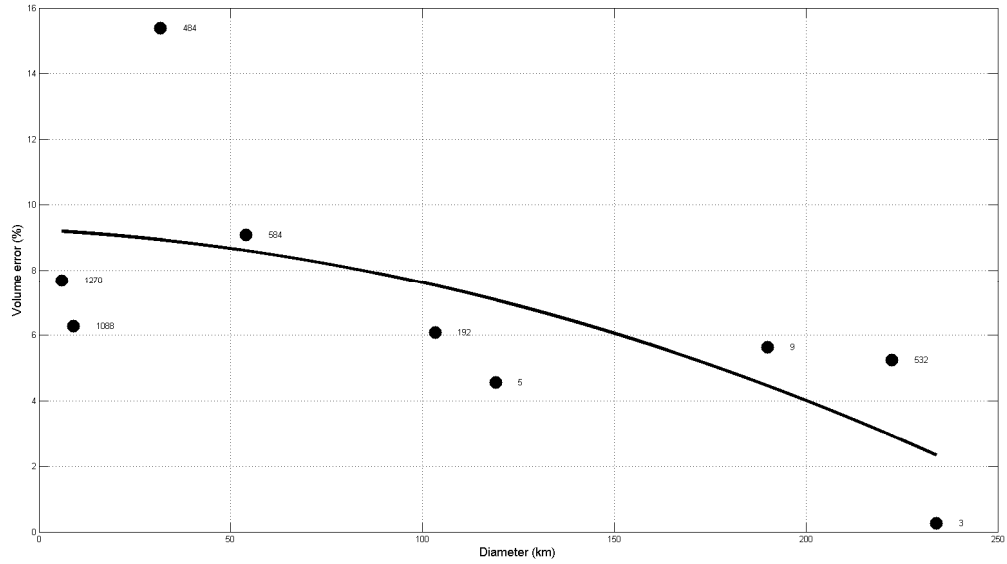

**Figure 5** – The volume error, respect to the complex model, in the geometrical scattering case vs. the asteroid effective diameter. The solid line is the best-fit parabola. The asteroids are identified with their progressive number.